# RESPONSIBLE DEEP LEARNING FOR SOFTWARE AS A MEDICAL DEVICE

Pratik Shah*. Ph.D.
UCI
(Pratik.Shah@uci.edu)

Jenna Lester. MD.
UCSF
(Jenna.Lester@ucsf.edu)

Jana G. Delfino. Ph.D.
Food and Drug Administration
(Jana.Delfino@fda.hhs.gov)

Vinay Pai. Ph.D. MBA
Food and Drug Administration
(Vinay.Pai@fda.hhs.gov)

Tools, models and statistical methods for signal processing and medical image analysis and training deep learning models to create research prototypes for eventual clinical applications are of special interest to the biomedical imaging community. But material and optical properties of biological tissues are complex and not easily captured by imaging devices [1]. Added complexity can be introduced by datasets with underrepresentation of medical images from races and ethnicities for deep learning, and limited knowledge about regulatory framework needed for commercialization and safety of emerging Artificial Intelligence (AI) and Machine Learning (ML) technologies for medical image analysis. This extended version of the workshop paper presented at the special session of the *2022 IEEE 19th International Symposium on Biomedical Imaging*, describes strategy and opportunities by University of California professors engaged in machine learning (section I) and clinical research (section II), and officials at the US FDA in Center for Devices & Radiological Health (CDRH) section IV, and the Office of Science and Engineering Laboratories (OSEL) section III. Performance evaluations of AI/ML models of skin (RGB), tissue biopsy (digital pathology), and lungs and kidneys (Magnetic Resonance, X-ray, Computed Tomography) medical images for regulatory evaluations and real-world deployment are discussed.

**I. DR. PRATIK SHAH: REGULATION WITH UNCERTAINTY QUANTIFIED DEEP LEARNING.** Strategies and methods for training of unbiased, high-performance locked and adaptive deep learning models with uncertainty quantifications. Special emphasis is made on the integration of unbiased imaging data and its positive impact by engendering fairness in deep learning model performance. The visually distinct anatomical features (e.g., eye, skin and prostate) require tailored imaging devices and high-fidelity capture of diseased and healthy tissues and their structure at the micro-and-macroscopic level for training deep learning models for clinical assessments. The biological reasons for diverse appearance of humans, such as blood perfusion in skin and pigmentation, cellular structure and tissue anatomy of prostate, and the retinal vasculature in the eyes, etc. provide naturally occurring targets for ionizing and non-ionizing radiation, autofluorescence, and chemical dyes for imaging abnormalities [1]. Recent advances in uncertainty-quantified deep learning architectures and statistical tools can be used for investigating the performance of the image-acquiring medical devices by estimating the pixel-level quality of acquired images prior to segmentation by deep learning models. For example, Dr. Shah's research group communicated an open-source deep learning toolkit that can perform end-to-end optimization of binary tumor segmentations by two different models from three different types of medical images of skin, microscopic pathology and MRI [2]. In other studies, they have also communicated model performance evaluations using concepts of reproducibility and generalizability by statistical inference and visual interpretability of medical images[3], [4]. This workshop paper proposes a machine learning framework for concepts of uncertainty and regulation to be embedded in the fields of medical imaging for deep learning. The key concept of uncertainty is often manifested as under specification, over or underfitting on training data that can limit model generalization to more diverse real-world data [5]. Lack of uncertainty estimation is often a major challenge precluding fully autonomous deployment of deep learning models for reproducible clinical interference from medical data [6]. Proposed solutions to evaluate model and data uncertainty include; a) extended intelligence: the blending of deep learning models with human intelligence to extend (not replace) physician capability (e.g. automatic tumor segmentation for medical image diagnosis) when necessary to improve model performance, b) intelligence augmentation (human-in-the-loop): a form of human-centered deep learning for augmenting (not replace) cognitive tasks (e.g. data recall) of physicians for clinical inference; c) ethical/responsible AI: model interpretability methods for why and how predictions were made, and whether they were fair towards underrepresented classes of patients and data subtypes; and d) regulatory science: application of the scientific method to improve the development, review, and oversight of new biologics, software and devices that require regulatory approvals. Generalized roadmap [**Fig. 1**] for real-world performance evaluation of equitable and fair deep learning for medical imaging should include robust research methods for: a) validation of data for desired inferences by causality estimates; b) use of fair machine learning tools for interpretability and explanation of model inferences; c) clinical validation using software as medical device digital health framework at the US FDA; c) policy framework for inclusivity of race and ethnicity labels and, d) involvement and education of clinicians, industry and data practitioners with fast evolving

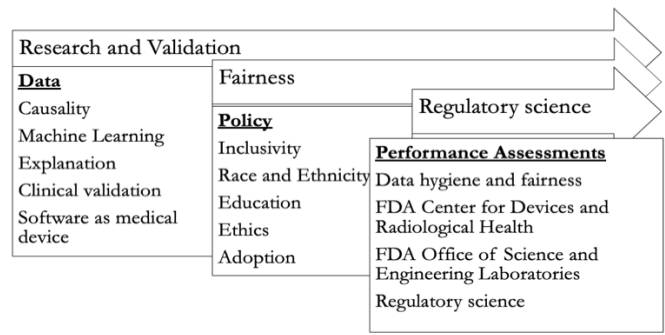

**Figure 1**: Roadmap for clinical deployment of deep learning models

*Correspondence: Pratik.Shah@uci.edu

fields of AI research and regulatory science policy framework [7].

## II. DR. JENNA LESTER: DERMATOGY IMAGING FOR FAIR MACHINE LEARNING

The skin of color program in dermatology research and education that aims to; a) improve care of patients with dark skin tones' b) promote research that allows diverse patients to achieve best dermatologic care possible; and c) develops education that enables providers to deliver the best care possible regardless of patient skin tone. Data inequity in dermatology education and research images with only 30% photos were of skin of color (brown to dark brown skin) are seen [8]. Research datasets are often private and difficult to assess diversity of available medical images. Many deep learning models are trained on datasets that do not contain a diverse range of images, in particular, they lack photos of dark skin and economically and socially marginalized individuals [9]. The dearth of images and data for skin of color is a known problem in dermatology, and can also be seen in imaging based diagnosis for other diseases, and can have a profound impact on the diagnostic reasoning of clinicians, health outcomes of patients, deep learning models [2]. For example, in 2014, prevalence of psoriasis was 3.6% among Caucasian adults; 1.9% among African Americans; and 1.6% among Hispanics [10]. In a Caucasian patient, psoriasis can be easily identified because it appears as a salmon color vs. in darker pigmentation of the skin, where it is not nearly as clear of a diagnosis [**Fig. 2**]. The darker pigmentation almost covers the manifestation of erythema, which is most often associated with psoriasis. Clinical studies have demonstrated that psoriasis plaques in African Americans can have a gray or violet shade to them which can make a psoriasis diagnosis challenging [10]. Psoriasis plaques can be difficult to distinguish from other skin conditions that are prevalent among skin of color patients such as lichen planus, discoid lupus, and sarcoidosis [10]. Dermatologists are likely to need a biopsy in skin of color of patients in order to make the differential diagnosis [11]. Expanding the clinical features of medical images to include balanced datasets with different races, ethnicities, and skin tones is a rapidly emerging field with promise to train fair deep learning systems [12]. Extrapolations for benchmarking algorithm performance on clinical images from socio-economic labels (income, zip codes, housing) can also be collected by medical imaging researchers for equitable deep learning. Inequity could lead to clinical diagnosis experiencing; a) anchoring bias: prematurely deciding on a diagnosis and not changing it with additional information; b) ascertainment bias: thinking is shaped by prior expectation; c) availability bias: judging things as being more likely or frequently occurring if they readily come to mind; d) confirmation bias: tendency to look for confirming evidence to support a diagnosis rather than look for disconfirming evidence to refute it; e) premature closure: accepting a diagnosis before it has been fully verified; f) representativeness restraint: drives diagnostician toward looking for prototypical manifestations of disease therefore missing atypical variants [13]. Solutions to address imbalance in inclusion and availability of diverse skin images include; a) prioritizing developing and testing algorithms on datasets with a diverse range of skin tones; b) sharing de-identified and patient consented data via publicly available repositories such as International Skin Imaging Collaboration (ISIC); c) addressing ethical concerns such as did the patients provide explicit consent for research, clinical and educational use of their images; and d)expanding medical education curriculum with specific examples of skin of color images and possible misdiagnosis [11].

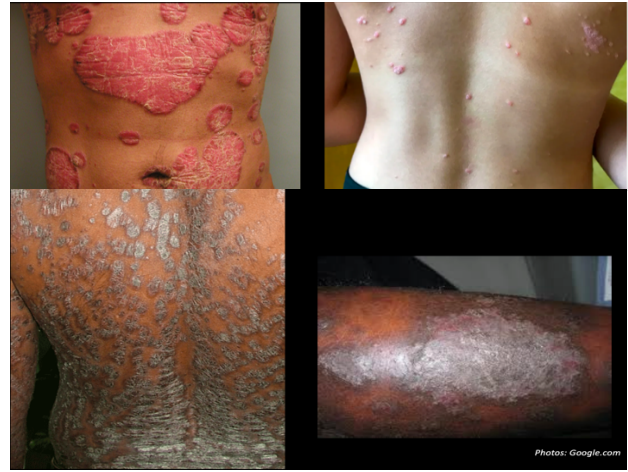

**Figure 2**: Misdiagnosis of Psoriasis in Caucasian (top panel) vs. skin of color patients (bottom panel).

## III. DR. JANA DELFINO: REGULATORY PERSPECTIVE ON AI/ML MEDICAL DEVICES

The U.S. FDA is working to establish a regulatory framework for software-based medical devices powered by deep learning. The FDA is working to help ensure patients and their caregivers continue to have access to safe and effective products powered by deep learning. Challenges faced by medical devices for computer aided diagnostics and quantitative imaging that incorporate AI/ML technologies are being researched by the OSEL within the FDA's Center for CDRH [14]. Key considerations include; a) regulatory landscape is continuing to evolve for AI/ML technology; b) devices that incorporate AI/ML technology are now being deployed; c) FDA recognizes that these devices present unique challenges and opportunities. The FDA OSEL conducts laboratory-based regulatory research to facilitate development and innovation of safe and effective medical devices and radiation emitting products. Key OSEL responsibilities include; a) provide scientific and engineering expertise, data, and analyses to support regulatory processes; b) Questions regarding both the training and testing datasets are important for fair machine learning model. Unfairness can be introduced at various stages in algorithm design, training, and testing from data such as selected human subjects are not; a) representative of the target population; b) do not include a complete spectrum of the target population; c) only few undergo the reference standard test; d) only some undergo one reference test, and others undergo another reference test; and e) the use of automation as a heuristic replacement for vigilant information seeking and processing. The OSEL AI/ML regulatory science research program has identified regulatory science gaps in lack of; a) methods that can enhance AI/ML algorithm training for clinical datasets that are typically much smaller than non-clinical datasets; b) clear definition or understanding of artifacts, limitations, and failure modes for fast-growing applications of Deep-Learning algorithms in the denoising and reconstruction of medical images; c) clear reference standard for assessing accuracy of AI/ML-based quantitative imaging (QI) and radiomics tools; d) assessment techniques to evaluate the trustworthiness of adaptive and autonomous AI/ML devices (for example, continuously learning algorithms; d) systematic approaches to address the robustness of various AI/ML input factors, such as data acquisition factors, patient demographics, and disease factors, to patient outcomes in a regulatory submission [**Fig. 3A**]. The OSEL AI/ML program activities are investigating approaches to overcome some of these challenges by; a) data augmentation, transferring learning, and other novel approaches to enhance AI/ML training/testing for small clinical datasets; b) study design and analysis methods for AI/ML- based computer-aided triage; d) non-clinical phantoms and test methods for assessing specific imaging performance claims for DL-based denoising and image reconstruction algorithms; e) imaging phantoms and computational models to support QI and radiomics assessment; f) assessment techniques for evaluating the reliability of adaptive AI/ML algorithms to support non-clinical test method development; and j) assessment approaches to estimate and report the robustness of AI/ML to variation in data acquisition factors [15].

**IV. DR. VINAY PAI: REGULATORY PERSPECTIVE ON REAL-WORLD PERFORMANCE OF DEEP LEARNING**

The goal of FDA Digital Health Center of Excellence launched in 2020 is to empower stakeholders to advance health care by fostering responsible and high-quality digital health innovation that meets standards of safety and effectiveness. A non-exhaustive currently marketed AI/ML-enabled medical devices public resource (https://www.fda.gov/medical-devices/digital-health-center-excellence) shows how AI/ML is being used across medical disciplines. Collaborative efforts with patient groups, healthcare providers, academics, and industry have been launched to help determine thresholds and performance evaluations for the metrics most critical to the real-world performance (RWP) of AI/ML-enabled medical devices[16]. The FDA's Software Precertification (Pre-Cert) Pilot Program and AI/ML-based Software as a Medical Device (SaMD) action plan have suggested an outline for the development of a possible future regulatory model for regulatory oversight of software-based medical devices and their real-world performance [3]. The strength of AI/ML systems is their ability to learn from real-world data and improve performance over time. Collection and monitoring of real-world data will support a total product lifecycle approach to the oversight of AI/ML-enabled software. Medical software manufacturers are encouraged to leverage the software technology's capability to capture real world performance data to understand user interactions with the SaMD, and to conduct ongoing monitoring of analytical and technical performance to support future intended uses. By gathering data on real-world use and performance of software, manufacturers can Improve their understanding of how their products are being used, identify opportunities for improvements, and respond proactively to safety or usability concerns. Software pre-specifications describes what aspects the manufacturer intends to change through learning and the algorithm change protocol explains how the algorithm will learn and change while remaining safe and effective. Key considerations include; a) what type of reference data are appropriate to utilize in measuring the performance of AI/ML software devices in the field; b) how much of the oversight should be performed by each stakeholder; c) how much data should be provided to the agency, and how often; d) how can the algorithms, models, and claims be validated and tested; e) how can feedback from end-users be incorporated into the training and evaluation of AI/ML-enabled software. Good machine learning practice for SaMD involves; a) multi-disciplinary expertise are leveraged throughout the total product life cycle; b) clinical study participants and data sets are representative of the intended population; c) training data Sets are independent of test sets; d) model design is tailored to the available data and reflects the intended use of the device; e) focus is placed on the performance of the human-AI team; f) testing demonstrates device performance during clinically relevant conditions; g) users are provided clear, essential information; and I) deployed models are monitored for performance and re-training risks are managed [**Fig. 3B**].

**4. REPORTING**

The authors report no conflicts or disclosures, and are in compliance with all ethical standards.

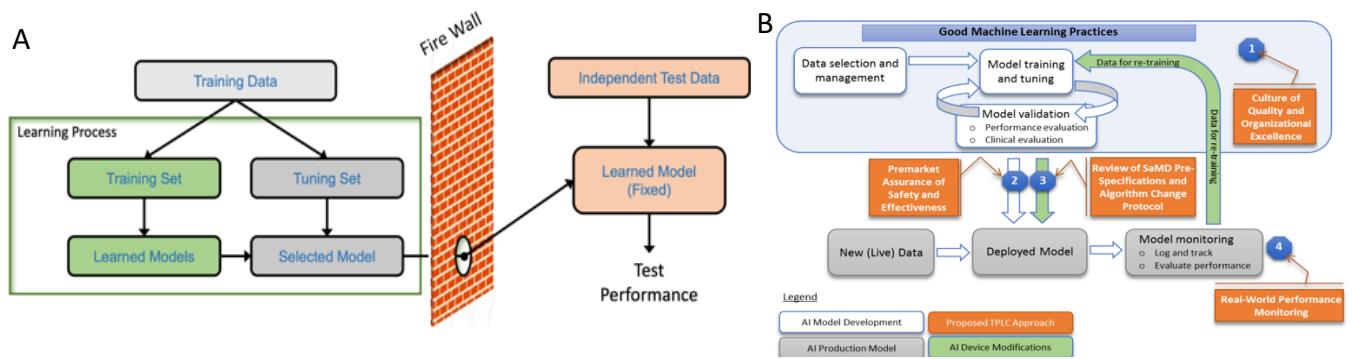

**Figure 3**: Proposed regulatory framework for Artificial Intelligence/Machine learning (AI/ML) based Software as Medical Device (SaMD). A. Proposed performance testing of AI/ML models with independent and balanced data. B. Proposed real-world performance monitoring of deployed AI/ML models under the FDA pre-cert framework.

## 5. REFERENCES

[1] P. Shah *et al.*, "Technology-enabled examinations of cardiac rhythm, optic nerve, oral health, tympanic membrane, gait and coordination evaluated jointly with routine health screenings: an observational study at the 2015 Kumbh Mela in India," *BMJ Open*, vol. 8, no. 4, p. e018774, Apr. 2018, doi: 10.1136/bmjopen-2017-018774.

[2] S. Ghosal and P. Shah, "A deep-learning toolkit for visualization and interpretation of segmented medical images," *Cell Reports Methods*, vol. 1, no. 7, p. 100107, Nov. 2021, doi: 10.1016/j.crmeth.2021.100107.

[3] A. Rana *et al.*, "Use of Deep Learning to Develop and Analyze Computational Hematoxylin and Eosin Staining of Prostate Core Biopsy Images for Tumor Diagnosis," *JAMA Network Open*, vol. 3, no. 5, p. e205111, May 2020, doi: 10.1001/jamanetworkopen.2020.5111.

[4] A. Bayat, C. Anderson, and P. Shah, "Automated end-to-end deep learning framework for classification and tumor localization from native non-stained pathology images," *SPIE*, vol. 2021, no. 6, doi: 10.1117/12.2582303.

[5] S. Ghosal, A. Xie, and P. Shah, "Uncertainty Quantified Deep Learning for Predicting Dice Coefficient of Digital Histopathology Image Segmentation," *arXiv preprint arXiv:2109.00115*, 2021.

[6] D. S. Bitterman, H. J. W. L. Aerts, and R. H. Mak, "Approaching autonomy in medical artificial intelligence," *The Lancet Digital Health*, vol. 2, no. 9, pp. e447–e449, Sep. 2020, doi: 10.1016/S2589-7500(20)30187-4.

[7] P. Shah *et al.*, "Artificial intelligence and machine learning in clinical development: a translational perspective," *npj Digital Medicine*, vol. 2, no. 1, pp. 1–5, Jul. 2019, doi: 10.1038/s41746-019-0148-3.

[8] J. C. Lester, J. L. Jia, L. Zhang, G. A. Okoye, and E. Linos, "Absence of images of skin of colour in publications of COVID-19 skin manifestations," *Br J Dermatol*, vol. 183, no. 3, pp. 593–595, Sep. 2020, doi: 10.1111/bjd.19258.

[9] G. A. Tadesse *et al.*, "Skin Tone Analysis for Representation in Educational Materials (STAR-ED) using machine learning," *NPJ Digit Med*, vol. 6, no. 1, p. 151, Aug. 2023, doi: 10.1038/s41746-023-00881-0.

[10] A. F. Alexis and P. Blackcloud, "Psoriasis in skin of color: epidemiology, genetics, clinical presentation, and treatment nuances," *J Clin Aesthet Dermatol*, vol. 7, no. 11, pp. 16–24, Nov. 2014.

[11] A. Fenton *et al.*, "Medical students' ability to diagnose common dermatologic conditions in skin of color," *J Am Acad Dermatol*, vol. 83, no. 3, pp. 957–958, Sep. 2020, doi: 10.1016/j.jaad.2019.12.078.

[12] A. Moiin, Ed., *Atlas of black skin*. Cham, Switzerland: Springer, 2020.

[13] P. Croskerry, "The importance of cognitive errors in diagnosis and strategies to minimize them," *Acad Med*, vol. 78, no. 8, pp. 775–780, Aug. 2003, doi: 10.1097/00001888-200308000-00003.

[14] N. Petrick *et al.*, "Regulatory considerations for medical imaging AI/ML devices in the United States: concepts and challenges," *J. Med. Imag.*, vol. 10, no. 05, Jun. 2023, doi: 10.1117/1.JMI.10.5.051804.

[15] B. J. Nelson, R. Zeng, M. B. K. Sammer, D. P. Frush, and J. G. Delfino, "An FDA Guide on Indications for Use and Device Reporting of Artificial Intelligence-Enabled Devices: Significance for Pediatric Use," *Journal of the American College of Radiology*, vol. 20, no. 8, pp. 738–741, Aug. 2023, doi: 10.1016/j.jacr.2023.06.004.

[16] M. Kunst *et al.*, "Real-World Performance of Large Vessel Occlusion Artificial Intelligence–Based Computer-Aided Triage and Notification Algorithms—What the Stroke Team Needs to Know," *Journal of the American College of Radiology*, p. S1546144023003356, May 2023, doi: 10.1016/j.jacr.2023.04.003.

## 6. AUTHOR BIOGRAPHIES

Dr. Pratik Shah Ph.D. is a faculty member at the University of California, Irvine. He leads a research lab that works on novel medical imaging, machine learning, and biological technologies to improve clinical, regulatory science and health outcomes in patients. Recent work from his lab has been published in Nature Digital Medicine, Cell press, Journal of American Medical Association, IEEE conferences. Pratik has BS, MS, and PhD degrees in biological sciences and completed fellowship training at Massachusetts General Hospital, the Broad Institute of MIT and Harvard, and Harvard Medical School.

Dr. Jenna Lester MD is a dermatologist and assistant professor in the Department of Dermatology at the School of Medicine in the University of California, San Francisco. Her clinic looks to address health disparities by providing dermatological care to people of color. Dr. Lester received her undergraduate degree at Harvard college and medical training from the Warren Alpert Medical School of Brown University. Her research has been published in JAMA Dermatology, The JAMA Network , Medscape and Journal of The American Academy of Dermatology.

Dr. Jana G. Delfino, PhD, is the Deputy Director for Medical Imaging and Digital Health in FDA's Division of Diagnostics, Imaging, and Software Reliability in the Office of Science and Engineering Laboratories. Dr. Delfino received her BS in Agricultural Engineering from the University of California at Davis and her PhD in Biomedical Engineering jointly from the Georgia Institute of Technology and Emory University Dr. Delfino has served in various roles at the FDA within both the Center for Devices and Radiological Health and the Center for Drug Evaluation and Research.

Dr. Vinay Pai, PhD, is a general Engineer in FDA's Center for Devices and Radiological Health, Office of Strategic Partnerships and Technology Innovation, Division of Digital Health (DDH),. His focus is on real-world performance of digital health devices, especially within the context of FDA's precertification pilot program. Before NIH, Vinay was a faculty at New York University School of Medicine developing techniques using proton and hyperpolarized helium magnetic resonance imaging to study cardiac and lung function. Vinay has a PhD in mechanical engineering from Florida State University, and an MBA from Johns Hopkins University.